\documentclass[12pt,letterpaper,oneside]{article}
\usepackage[numbers]{natbib}
\usepackage{graphicx}
\usepackage{multicol}
\usepackage{epsfig}
\usepackage{caption}

\def\nh{$n_{\mathrm{H}}$\/}

\def\nc{$N_{\rm c}$\/}
\def\rfe{$R_{\rm FeII}$}

\def\feiiq{\rm Fe{\sc ii}$\lambda$4570\/}

\def\ltsima{$\; \buildrel < \over \sim \;$}
\def\ltsim{\lower.5ex\hbox{\ltsima}}  
\def\gtsima{$\; \buildrel > \over \sim \;$}

\def\gtsim{\lower.5ex\hbox{\gtsima}}

\def\ha{{\sc H}$\alpha$}
\def\lya{{ Ly}$\alpha$}
\def\civ{{\sc{Civ}}$\lambda$1549\/}

\def\cmq{cm$^{-2}$\/}
\def\cm3{cm$^{-3}$\/}
\def\hb{{\sc{H}}$\beta$\/}

\def\mgii{{Mg\sc{ii}}$\lambda$2800\/}

\def\o4363{{\sc{[Oiii]}}$\lambda$4363\/}

\def\oi{{\sc{Oi}}$\lambda$8446\/}
\def\caii{{Ca{\sc ii}}}

\def\feiiopt{{Fe \sc{ii}}$_{\rm opt}$\/}
\def\feii{{Fe\sc{ii}}\/}

\def\fe{{\sc{Fe}}\/}

\def\fe76087{{\sc [Fe vii]}$\lambda$6087\/}

\def\kms{km~s$^{-1}$}

\def\hii{H{\sc ii}\/}

\def\apj{ApJ}
\def\apjl{ApJL}
\def\apjs{ApJS}

\def\mnras{MNRAS}
\def\aap{AAp}

\usepackage{titlesec}
\titleformat{\section}
  {\normalfont\scshape}{\thesection}{1em}{}
\title{\bf Low Ionization Emission Lines in Quasars\\ {\small \bf Clues  from O{\sc i} 8446 and the \caii\ Triplet}\footnote{Submitted to the {\em Astronomical Review}.}}
\author{Paola Marziani\footnote{INAF, Osservatorio Astronomico di Padova, Italia.},  Mary Loli Mart\'{\i}nez-Aldama, Deborah Dultzin,\footnote{Instituto de  Astronom\'{\i}a, UNAM, M\'exico.}\\ and  Jack W. Sulentic\footnote{Instituto de  Astrof\'{\i}sica  de Andaluc\'{\i}a (CSIC), Granada, Spain.} }
\date{}
\begin{document}
\maketitle

\begin{abstract}
The formation of low emission lines in quasars and active galactic nuclei is
still an open issue. Aided by the organizing power of the 4D eigenvector 1 scheme,  we review  basic developments since  the 1980s, devoting  special attention to 
the \caii\ IR triplet and the O{\sc i} 8446  emission lines. Coverage of these lines
is cumbersome since they are shifted in an inconvenient IR domain already at
modest redshifts ($\approx$ 0.2). Their detection is also difficult since they are
faint and often   buried in the \caii\ absorption of the host galaxy.   We
discuss  how these lines can provide unambiguous constraints on the physical
conditions of the broad line emitting regions of quasars when detected in
emission, and summarise preliminary results for a sample of luminous,
intermediate redshift quasars.
\end{abstract}

\section{Introduction}

Several fields of quasar astronomy are still on uncertain grounds. Although a basic understanding in terms of accretion on a supermassive black hole seems now established \citep[][and references therein]{donofrioetal12}, there are several aspects of quasars that still defy comprehension or  at least a consistent view. Even if we   learned many important things on quasars in the past fifty years,  we have to acknowledge that  mechanisms giving rise to continuum emission in several bands of the electromagnetic spectrum are still poorly understood, as recently stressed by several authors \citep{donofrioetal12,antonucci13}.  Specifically, it is not  clear how soft X-ray emission originates, and why radio emission is strong in a minority of quasar and faint in the wide majority. And we do not know well how quasars evolve since there is no major sign of evolution from redshift almost 0 to (at least) 4, when the Universe was about $\frac{1}{4}$\ of its current age. A maze of apparently confusing pieces of evidence still fogs the connection between evolution and unification schemes.   Not to mention the processes that give rise to prominent line emission. Going beyond heuristic considerations based on the nebular physics  developed to interpret \hii\ regions and planetary nebul\ae\ in the 1950s and 1960s has proved a daunting challenge. The complexity of quasar emitting regions, dense gas exposed to an incredibly strong radiation field became clear relatively early after quasar discovery \citep{davidson72,davidsonnetzer79}. So it is perhaps not surprising if some basic and important issues have been left in a limbo, without a clear understanding ever being reached, as discussed by  \citet{sulenticetal12a}. One of the issues concerns the emission mechanism of low-ionization lines in quasars.

If we start looking at quasars from their optical and UV spectra, one of their notable properties is the coexistence of low and high emission lines, where for low and high ionization we understand lines emitted by ionic species whose ionization potential is $\ltsim 20$ eV (hydrogen, singly ionized ionic species of magnesium, carbon, iron, calcium), and  $\gtsim 40$ eV (triply ionized carbon, helium, four times ionised nitrogen)  respectively. There is now convincing evidence that high ionization   and hydrogen lines  are mainly photoionized by the strong quasar continuum since emission line luminosity is proportional to continuum luminosity (at least to a first approximation, \citep{shuder81}) and lines respond to continuum luminosity changes \citep{petersonetal82}. The issue is still not  solved at the time of writing for  \feii, whose emission features extend from the far UV to the near IR and are in some cases dominating the appearance of optical quasar spectra \citep{marzianietal06,vandenberketal01}. 

\section{Organizing Quasar Diversity: Low Ionization Lines in the 4DE1 Context}

The 4D Eigenvector 1 (4DE1)  is a parameter space best suited to understand the phenomenology of the of the broad line region (BLR), nowadays an ``umbrella term'' that embraces an unresolved but complex region where all broad lines are emitted 
\citep{sulenticetal00a,sulenticetal00b,marzianietal01,marzianietal03a,marzianietal03b,sulenticetal07,zamfiretal08,negreteetal12}. Following the 4DE1 scheme, sources can be divided in two populations, A and B, with separation at FWHM(H$\beta$) = 4000 km s$^{-1}$. Several differences are found between sources above 
and below this limits (see Refs. \citep{sulenticetal07,sulenticetal11} for reviews). Population A shows: (1) a scarcity of  RL sources, (2) strong/moderate \feii\ emission, (3) a soft X-ray excess, (4) high-ionization broad lines  showing blueshift/asymmetry and (5) broad line profiles are best described by Lorentz fits.  Meanwhile,  Population B: (1) includes the large majority of the RL sources, (2) shows weak/moderate \feii\ emission,  (3) does not show a soft X-ray excess nor a prominent high ionization line (HIL) blueshift/asymmetry, and (4) their \hb\ profiles are  best fit with double Gaussian  models.  Considering that the eigenvector 1   identifies a single dimension in a multidimensional space, it is appropriate to talk of a ``sequence along the eigenvector 1'' (see for example Fig. 1 of \citet{marzianisulentic12a}). The equivalent width of  HILs is anti-correlated with  \feiiq\ equivalent width \citep{borosongreen92,sulenticetal00a},  suggesting a systematic decrease in ionization   level from  population B to population A \citep{marzianietal01}. In addition, there are systematic Eddington ratio differences along the sequence,  with Pop. A sources being  higher radiators.  The blue ward asymmetry observed in the \civ\ profile indicates that at least a part of the HILs is emitted in outflowing gas where the receding part of the flow is obscured \citep{gaskell82,marzianietal96,richardsetal06}. The prominence of the outflow is maximised at the high Eddington ratio end of the 4DE1 sequence   \citep{marzianietal96,marzianietal10}.   Interestingly enough, there are trends that are  preserved along the 4DE1 sequence. For example, \mgii\ and    \feii\ show systematically narrower profiles than \hb\ \citep{marzianietal13a,sulenticetal06}. If motion is predominantly virial, then \feii\ and \mgii\ might be emitted at larger distance from the ionizing sources than \hb.


In Figure 1 of \citet{dultzinhacyanetal99} we can see two extreme orientations.  At one extreme of this configuration -- when the central source is seen face-on, parallel to the axis of the cone --
\civ\ shows an asymmetric profile and a minimum equivalent width, and \feii\ -- \caii\ emission is strong. The accretion disk obscures  the receding half of the emitting clouds, allowing us to see only the approaching side. When the line of sight forms an angle with respect to the disk axis,  we expect  to have symmetric \civ\  profiles and less \feii\ -- \caii\ emission. The \hb\ core shares the same behavior as   \feiiopt. It is unshifted, symmetric, and is consistent with a flattened or nearly spherical distribution of clouds.  

Summing up, the location of a source  in the 4DE1 sequence appears to be governed by Eddington ratio  with orientation   acting as a source of scatter \citep{marzianietal01,sulenticetal03}.    Black hole mass  and, to a lower extent, metallicity are also found to have a broadening effect on the 4DE1 sequence \citep{sulenticetal01,zamanovmarziani02}. Even if 4DE1 eases the systematization of quasar spectral properties, a full understanding of some  parameters used in the formulation of the 4DE1 itself is still missing. 

\begin{figure}[t]
\centering
\includegraphics[width=6.in]{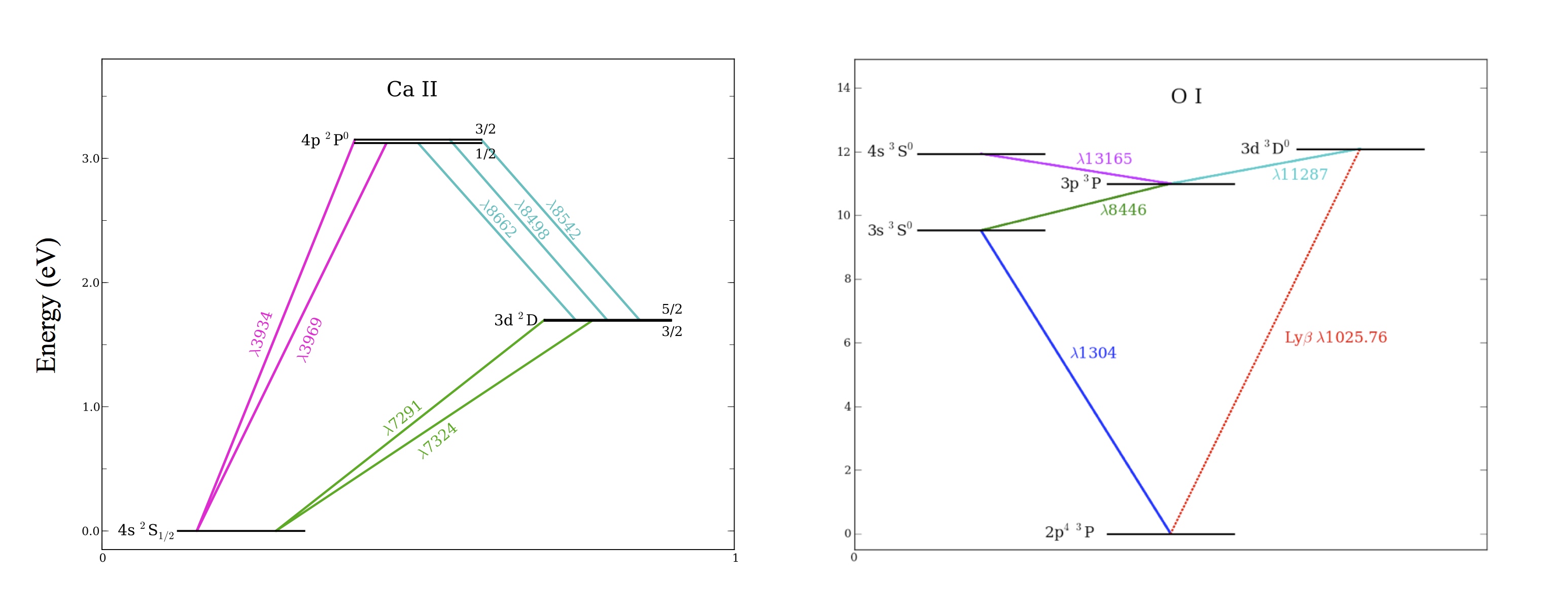}
\caption{Highly simplified Grotrian diagrams for  singly ionized calcium (left) and neutral oxygen (right). Only levels leading to main transitions discussed in the text are shown.  Note that   {Ca \sc ii} and  {O \sc i} are not on the same vertical scale. The red line identifies  the {O \sc i} transition that is affected by Bowen resonance with Ly$\beta$\ photons. }
\label{fig:grotrian}
\end{figure}


\section{Low Ionization Lines: the \feii\ Problem}

Photoionization models have troubles in explaining the intensities of  some prominent low-ionization lines (LILs): the wording ``\feii\  problem'' has been widely used to address 
the issue of the impossibility to reproduce the strength of \feii\ (optical and UV)  with simple photoionization models \cite{willsetal85}.   The complexity of the energy level structure of the \feii\ ion and the uncertainties in the 
atomic data make theoretical models calculations very difficult \citep{netzerwills83,joly87,collinjoly00}.  
It was recognised in the late 1970s that an important -- albeit insufficient -- mechanism for producing   optical \feii\ emission is    collisional excitation \cite{phillips78b}.   Over the years, the importance of continuum and line fluorescence (especially  \lya\ and \feii\ lines self fluorescence) has been also recognized \citep{sigutpradhan03} for explaining the whole \feii\  emission from UV to IR. However, the  large number of electronic transitions of the  singly ionized iron ion  form a pseudo-continuum especially prominent in the UV range between 2000 and 4000 \AA.  As stressed since long \citep{willsetal85},   \feii\ can be the single largest contributor to the emission line spectra, and is therefore a major cooling agent in the BLR. The effect of cooling is to lower excitation temperature and hence to saturate \feii\ emission  with increasing iron abundance   \cite{dumontetal98,sigutpradhan03,verneretal99}.  Photoionization models fail to account for strong  \feii\ emitters with \rfe $\gtsim$ 1, where \rfe\ in the intensity ratio between the \feii\ blend at 4570 \AA\ and \hb.   It is still unclear whether models are inadequate, or an  independent heating source is needed.  At any rate, the problems of photoionization models  led to the suggestion that  \feii\ is emitted in a region which is not radiatively heated, arguably of  high density  (\nh $\sim$10$^{11.5}$ --10$^{12}$ cm$^{-3}$), high column density (\nc\ $\sim$ 10$^{24}$ cm$^{-2}$), and relatively low electron temperature, T $\sim$ 8000 K\ \citep{collinsouffrinetal80,collinsouffrinetal86,joly87,ferlandpersson89,matsuokaetal07}. A possible scenario involves a shocked region shielded from the central continuum source \citep{jolyetal08}. 

The locus of the emission of  all LILs  in the BLR is subject of debate since many years. As far as spatial location of \feii\ emission is concerned,  circumstantial evidence indicated the outer 
part of the BLR \citep{rodriguezardilaetal02a} or  the outer part of the accretion disk \citep{collinsouffrinetal88,dultzinhacyanetal99}. Recent efforts were successful in estimating the reverberation distance of \feiiopt\ in two low-$z$ sources with \feiiopt\ of moderate intensity \citep{barthetal13}, indicating a time-delay distance $\approx$ 1.5 times larger than the one of \hb, as found for Mg{\sc ii}$\lambda$2800 \citep{marzianietal13}.

\section{The Calcium IR Triplet and \oi}

\paragraph {\oi\ --} Given  the difficult interpretation and prediction of the  \feii\ spectrum, one could resort to the study of the a much simpler ionic species. Especially helpful could be the near IR   \caii\ triplet at 8498, 8542, 8662 \AA\ (hereafter indicated with \caii\ for brevity), and the \oi\ emission lines. It has been suggested that these lines are emitted under the same physical  conditions as the \feii\ lines \citep{joly87,ferlandpersson89,rodriguezardilaetal02a,matsuokaetal07,matsuokaetal08}.

\citet{grandi80} carried out a landmark observational study of the \oi\ line 
for thirteen Seyfert 1 galaxies, showing that the strength and width of these lines are very 
similar to the H$\alpha$ line. He suggested that the emission is associated to the BLR   
due to the lack of a narrow component.\footnote{\citet{landtetal08} were able to detect the 
narrow component in Seyfert 1 and 1.5 galaxies, therefore showing that \oi\ emission is not a exclusive of 
the BLR. In the following we will be concerned only with broad \oi\ emission.} \citet{netzerpenston76} proposed  Bowen fluorescence  (or Ly$\beta$ fluorescence)   as the main excitation process for \oi. There is a   coincidence of the energy levels between hydrogen and neutral oxygen: 
Ly$\beta$ ($\lambda$1025.72 \AA) can excite the {O \sc i} ground state resonance transition 2p $^{3} P \rightarrow$ 3d $^3D^\mathrm{o}$  ($\lambda$1025.88 \AA) if the thermal line width is as expected at $T \sim10^4$ K, a typical temperature of photoionized gases. A simplified Grotrian diagram of the O$^0$\ atom is shown in Fig. \ref{fig:grotrian}; the red line identifies the \lya-induced transition.  The \oi\ 
line can also be excited by \ha\ if the gas is optically thick in this line \citep{grandi80}. During the 
cascade emission to the ground level, other {O \sc i} photons are emitted: O{\sc i} $\lambda$11287 and O{\sc i} $\lambda$1304. Hence, the photon flux ratio between $\lambda$11287 and $\lambda$8446 has to be exactly one; 
smaller values suggest that other excitation processes increase the emission of $\lambda$8446. 

The O{\sc i} $\lambda$7775 line was first detected by \citet{rodriguezardilaetal02b} in one object, and later detected for several objects by \citet{landtetal08}. This 
line is produced by collisional excitation,\footnote{The lower level associated to this feature, 3s $^5S^\mathrm{o}$, is the ground level of the quintuplet system of O{\sc i} and is metastable.} and thus the presence of this mechanism was 
corroborated.  The transitions probabilities imply  O{\sc i} $\lambda$7775 /$\lambda$ 8446 $\sim$ 1.1 for recombination, and O{\sc i} $\lambda$7775 /
$\lambda$8446 $\sim$ 0.3 for collisional excitation. \citet{landtetal08} could recognize both contributions in different proportions, given that the observed O{\sc i} $\lambda$8446  intensity is usually in excess with respect to value expected from Ly$\beta$\ fluorescence \citep{rodriguezardilaetal02b}. \citet{matsuokaetal07}, using photoionization models, verified that both  Ly$\beta$ fluorescence and collisional excitation produce the O{\sc i} emission in a gas with $n_\mathrm{H} \sim$10$^{11.5}$ cm$^{-3}$, illuminated by radiation with a ionization parameter $U \sim$10$^{-2.5}$. Therefore,  Bowen fluorescence is not the only excitation process for \oi\ although it accounts for most or at least half of the photon flux in the wide majority of the sources studied by \citet{landtetal08}. Continuum fluorescence,  recombination and collisional excitation by electrons  are likely be present  in different proportions for each source. 

\begin{figure}[t]
\centering
\includegraphics[width=4.in]{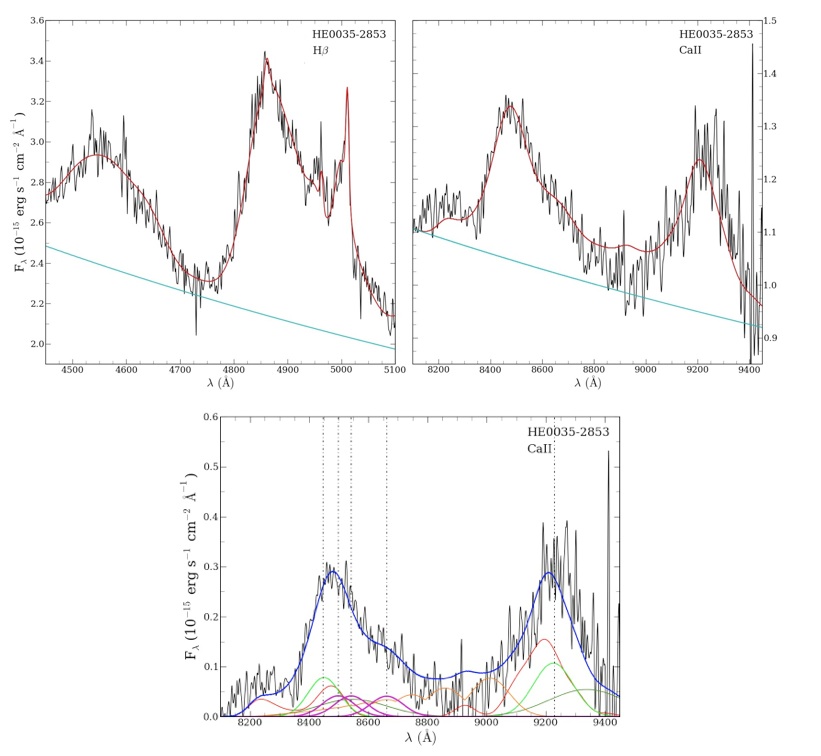}\\
\caption{Analysis of a Pop.  B quasar, HE 0035--2853. Top:  the left and right panels show the   \hb, and  the \caii\ and \oi\ spectral ranges respectively, in their rest frame.  Abscissa is rest frame wavelength and ordinate rest frame specific flux.  The spectra were obtained from VLT-ISAAC observations at different epochs. The red line traces the model spectrum including emission, absorption (negligible in this cases) components and continuum. The pale blue lines mark the modelled continuum underlying all emission features.  Bottom: various component included in the multicomponent fit carried out to measure the \oi\ (lemmon green + dark green VBC) and \caii\ (magenta) lines, again in rest frame specific flux versus rest frame wavelength.  This panel shows the continuum subtracted  \caii\ spectral range, with model spectrum colored  navy blue. The orange line traces high order Paschen lines, while the red line represents \feii\ emission according to the \citet{garciarissmannetal12} template.  The green lined at $\approx$ 9200 \AA\ represent Pa 9, with both broad (lemmon green) and very broad (dark green= component shown.}
\label{fig:analysis}
\end{figure}

\paragraph{\caii\ --} Singly-ionized calcium emission is mainly due to  three multiplets emerging from five levels (Fig. \ref{fig:grotrian}): (1) the optical 
H and K lines ($\lambda$3933, $\lambda$3968 \AA) are emitted from level 4p  to the ground 
level 4s; (2) the infrared multiplet,   $\lambda$8498, $\lambda$8542, $\lambda$8662 \AA\ arises from transitions from  the 4p level to  the 3d metastable levels; (3) finally, the forbidden multiplet, $\lambda$7291, $\lambda$7324 \AA, arises 
from the 3d metastable levels to the ground level. The ionization potential of Ca$^+$\ is 11.871 eV, so the gas where 
the IR triplet is emitted should be strongly shielded from incident ionization at wavelengths shortward of Ly$\alpha$ 
($\approx$10.20 eV). We expect that this region is  neutral, optically thick and with high 
column density \citep{ferlandpersson89,joly89}. Due to the similarity between the energy of Ly$\alpha$ \
and the ionization potential of Ca$^+$, the 3d metastable level is highly populated and the collisional excitation process leading to the \caii\ IR triplet emission is efficient. Attempting to reproduce the gas conditions  requires that  processes   like  free-free heating,  photoionization from excited levels,  bound-free heating and Compton recoil ionization be included in photoionization calculations. These processes are able to heat neutral gas with great depths \citep{ferlandpersson89} and produce low ionization lines without  the need to call upon nonradiative heating processes. 

The first survey  of \caii\ near-IR triplet in luminous Seyfert 1 galaxies and low-$z$\ quasars was performed by  \citet{persson88}. He could clearly  detect the \caii\ emission in fourteen objects, and  he  could also test  that the three lines of the calcium triplet emerge from an optically thick region since  they show the same intensity within the  uncertainties.  In addition,  \citet{persson88} found that the field velocities of \caii\ and \oi\ are closely related,   with  \oi\ being somewhat  narrower. On other hand, \citet{rodriguezardilaetal02a} did not find differences in widths, suggesting  that   \oi\ and \caii\   are emitted at the same distance from the ionizing source. 

Comparing Pa$\beta$ $\lambda$12818, O{\sc i} $\lambda$11287, \feii$\lambda$11127 and \caii\ triplet,  \citet{rodriguezardilaetal02a} proposed that Pa$\beta$ \ is emitted in an intermediate part of the BLR, while \feii, \oi\ and \caii\ are emitted in the outermost part of the BLR. \citet{matsuokaetal07} compared EW predicted from photoionization models to the observed ones, and concluded that HILs should be emitted in regions which are more highly ionized  and thus closer to the ionizing source. Comparing the intensity ratios of \caii/\oi\  and O{\sc i} 
$\lambda$11287/\oi\ with photoionization models predictions for several sources, 
 \citet{matsuokaetal08} found that these lines are being emitted in gases with similar density, but exposed to different ionizing photon flux. This is likely due to a difference in distance from the central continuum source  of the quasar. 
 
 A natural site of emission is the  accretion disk  that offers the high 
column density required for emitting LILs  \citep{collinsouffrin87,collinsouffrinetal88}.  \citet{dultzinhacyanetal99} found a very similar behavior between the ratios of \feiiq/\hb\ and \caii/H$\beta$ as a function of FWHM(H$\beta$), respectively, suggesting that they arise in the same region,  an  optically thick accretion disk. \citet{collinsouffrinetal86} and   \citet{ferlandpersson89} proposed that there may be a nonradiative source  of heating which can be the dissipation of mechanical heating inside the accretion disk.  We want to point out that although the emitting regions may be at the same distance, they do not have to be  the same physical region. For example, we may envision a configuration of plane parallel distribution of BLR clouds above and below the accretion disk.  In this configuration the clouds dynamics is dominated by gravity and thus the emitting gas is virialized. Some of these clouds might be at the same distance from the ionization source as some regions within the accretion disk. And thus they share the same dynamics. However, they do not necessarily share the same physical conditions. 
On the other hand,  the presence of non gravitational  forces  such as radiation pressure is not excluded \citep{netzermarziani10}. As we explain below,  the emitting gas within the BLR may be subject to different dynamics. This is why in order to accurately fit a broad line we often need more than one component \citep{marzianietal08,negreteetal12}.


\begin{figure}[t]
\centering
\includegraphics[width=7.75cm]{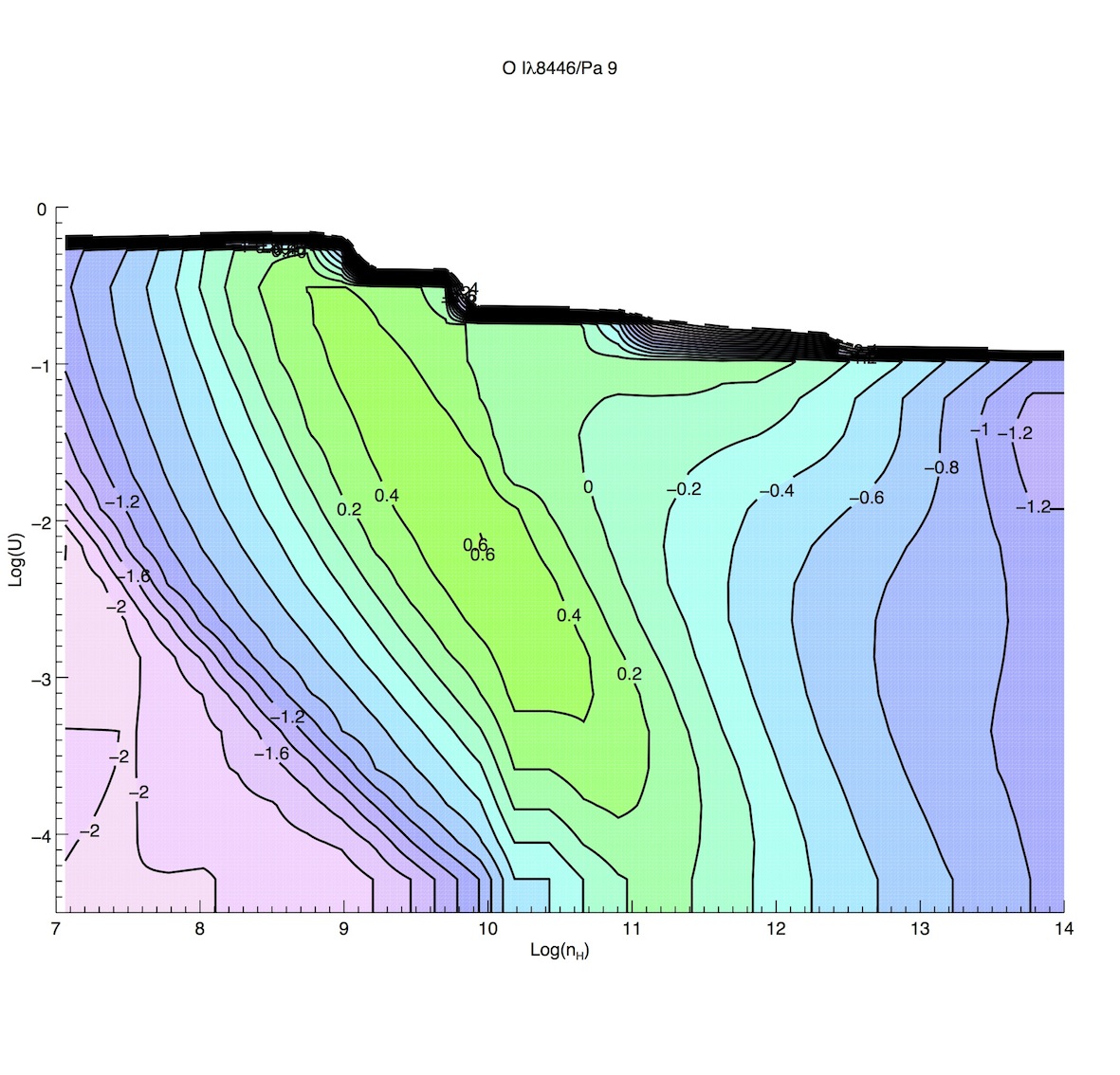}
\includegraphics[width=7.75cm]{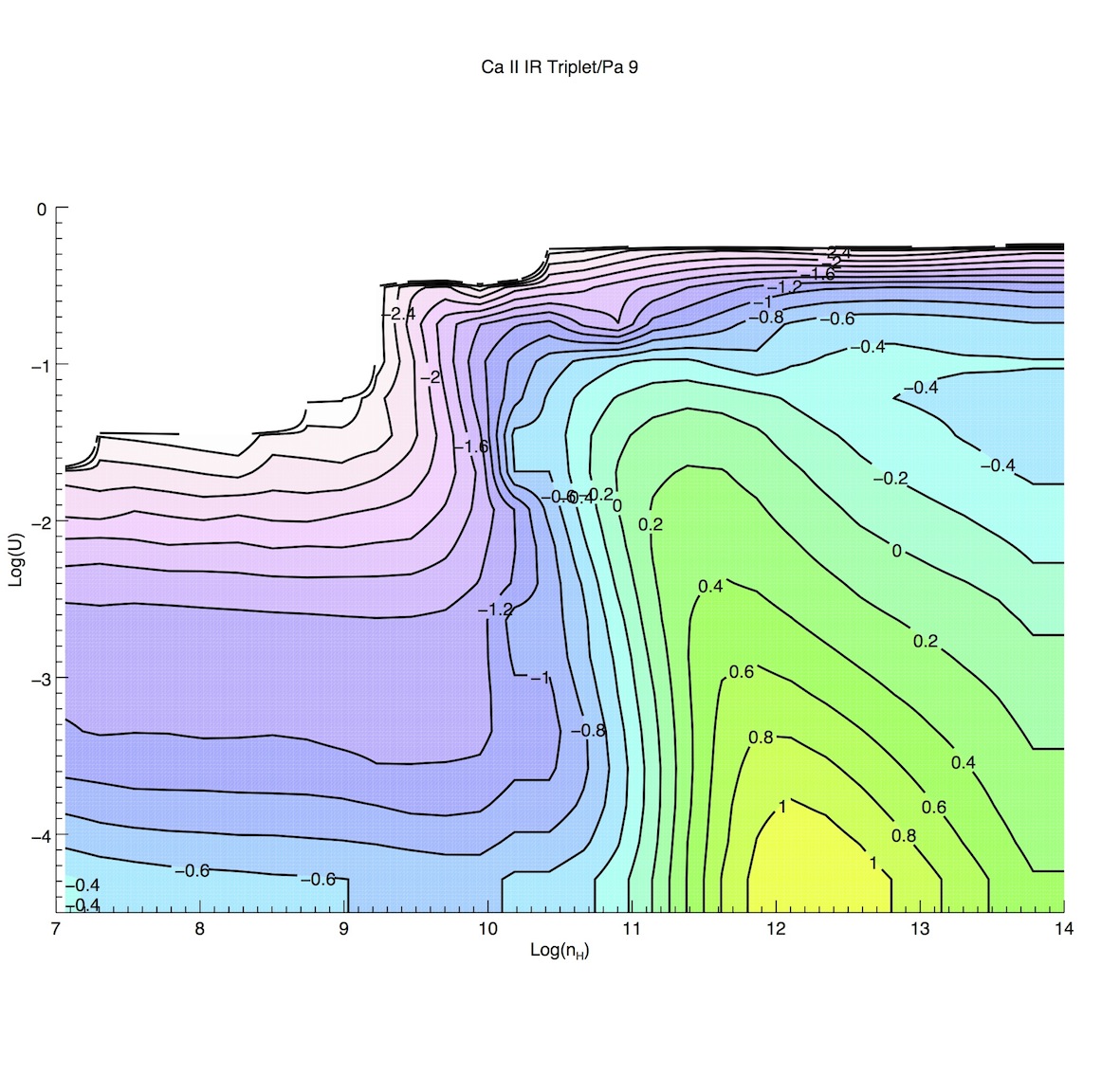}
\caption{Behavior of the  \oi/Pa9 and \caii/Pa9 intensity ratios as a function of logarithm of hydrogen density \nh\ and ionization parameter $U$. Isophotes are drawn at 0.2 dex intervals. The white areas at the top of both diagrams indicate \caii\ and \oi\ $\rightarrow$ 0. Qualitatively similar trends in the plane (\nh, $U$) are obtained if ratios are computed for \hb\  in place of Pa 9.}
\label{fig:nhu}
\end{figure}

\section{Preliminary Results from  \caii\ and \oi\ Observations of Quasars}

Several important aspects emerge from the previous discussion: LILs like \feii\ and \caii\ can be both emitted in  a low ionization, dense medium of high column density, possibly shielded from the central continuum sources. The narrower line width of \feii\ and \mgii\ indicate that the distance from the central continuum source could be larger than for other LILs like the Balmer lines. The 4DE1 sequence is defined as a sequence of increasing \feii\ prominence, and we have seen that \feii\ emission is still not satisfactorily modelled. While there has been a considerable progress with photoionization models of \feii, and photoionization is now supported by the observation of the \feii\ response to continuum changes \citep{barthetal13}, it is not clear if strong \feii\ emitters can be explained in a pure photoionization scheme. According to Collin and Joly, a different source of heating is needed. The strongest \feii\ emitters show however no solution of continuity with fainter \feii\ emitters in the 4DE1 sequence. By Occam's razor, it should be avoided to invoke an additional mechanism if no discontinuity is observed. As we have seen, the \caii\ and \oi\ line formation is better understood. What happens to these lines along the 4DE1 sequence? 

We will not present here a full answer to this question. There are several difficulties to face before a full answer can be obtained.  An important side of the issue is observational:  \caii\ and \oi\ are shifted in the near IR even if redshift is just $z \gtsim 0.2$. Obtaining near IR spectra with the same quality (S/N and dispersion) of optical ones has become possible only in recent times, and there is currently  a very limited number of observatories where these lines can be observed in quasars. The consequence is that there are almost no data available even for luminous, low- and intermediate redshift quasars that make most of the samples over which the eigenvector correlations are analysed. The very  spectral range around $1 \mu$m in rest frame is not well understood. The continuum shape is uncertain, since  at $1 \mu$m the low energy end of the continuum emitted by the accretion disk, and the high energy end of the hot dust dominating in the mid and far IR merge creating a ``hollow'' in the spectral energy distribution \citep{landtetal11}. In addition, the  stellar population of the underlying galaxy bulge is expected to emit a spectrum peaking at $\approx 1\mu$m, with \caii\ in absorption. Turning to the emission lines, significant \feii\  has been revealed in the range 8000 -- 10000 \AA\ \citep{sigutpradhan98,rodriguezardilaetal02a}.  It is interesting to note that the main mechanism  responsible for the near IR \feii\ emission is   fluorescence excitation by Ly$\alpha$  and in a minimal proportion collisional  excitation since the  lines at 8900 -- 9300 \AA\ are emitted by cascading from a high excitation level that can be populated only by \lya\ fluorescence. The intensity of these lines are predicted by photoionization calculations and it is possible to model them in the observed spectrum using a template developed {\em ad hoc} \citep{garciarissmannetal12}.  The last additional complication in that the \oi\ and \caii\ and high order Paschen lines lines are blended together.  Extracting information on line fluxes of \oi\ and \caii\ requires therefore a careful modelling of the continuum and a multicomponent fit to the emission blend. An example is shown in Fig. \ref{fig:analysis}. 

We analyzed a sample of 14 high luminosity Hamburg-ESO quasars  with $M_\mathrm{B} < -26$ and 0.85 $< z <$ 1.64 using the  VLT ISAAC IR spectrometer during 2010 in service mode.\footnote{A full account of   observations of observantions and results will be presented elsewhere, in Mart\'{\i}nez-Aldama et al., in preparation.} These sources had previous high quality \hb\ observations obtained with the same instrument  \citep{sulenticetal04,sulenticetal06,marzianietal09}. In addition, the Pa9 line recorded on the same spectra can be used as a proxy of \hb.  Our sample  contains 4 Pop. A and 10 Pop. B sources making lower intensity \feii\  sources  (0.1 $\ltsim$ \rfe $\ltsim 0.5$) well represented in our  sample.   

A major result is that \caii\ is detected in all sources, even in sources where \feii\ is relatively faint. This has profound implications. Fig. \ref{fig:nhu} shows the behavior of predicted   \oi/Pa9 and \caii/Pa9 intensity ratios as a function of hydrogen density \nh\ and ionization parameter $U$. The ratios have been computed from an array of {\sc cloudy 08}  \citep{ferlandetal13}  simulations that assumed constant \nh\ and $U$\ for a gas slab   of solar metallicity, \nc = 10$^{23}$ \cmq, illuminated by a standard quasar continuum.  Given the S/N ratio of our spectra, a detection of \caii\ implies a sharp lower limit to density, \nh $\approx$ 10$^{11.5}$ \cm3. Lower density values would make \caii\ undetectable. A second result is that the strongest \caii\ emitters in our sample,  with  \caii/\hb $\approx$ 0.35  imply  $ \log U \sim -2$, and \nh $\sim 10^{12}$ \cm3. Inferences on density and ionization parameters are reinforced by the measures of the \caii/\hb\ ratio in extreme Pop. A sources like I Zw 1 or Mark 231: \caii/\hb $\approx$ 0.4 -- 0.5 requires $-2.5 \ltsim \log U \ltsim -2$, and \nh $\sim 10^{12}$ \cm3. Therefore,  \caii\ appears to be accounted for within the framework of photoionization models. An additional result emerging from the simulations is the different behaviour of \oi\ and \caii\ in the plane (\nh, $U$):   \oi\ emission if favoured at higher photon flux than \caii. We interpret the difference in behaviour as due to the Bowen fluorescence mechanism that is strongly influenced by the ionizing photon flux. The Bowen mechanism has indeed been found to be the major contributor to the \oi\ intensity in  most active nuclei studied by \citet{landtetal08}. The important implication is that   \oi\ emission  can originate in deep regions exposed to a large ionizing photon flux provided that column density is high, \nc $\gtsim$ 10$^{23}$ \cmq. 

The \hb\ line profile close to the line base  is significantly redshifted in the spectra of Pop. B sources; however, the ratio shift/width $\approx$ 0.1 -- 0.2 is modest and indicates predominance  of virial broadening. If so, the emitting gas should be exposed to the strongest ionizing photon flux since it is at the smallest distance from the ionizing source \citep{koristagoad04}. Empirically, a model with 2 Gaussian functions has been highly successful in reproducing the \hb\ profiles of Pop. B sources \citep{marzianietal03b,marzianietal09,zamfiretal10}. The broadest Gaussian has been aptly named  ``very broad component,'' with FWHM $\sim 10000$ \kms. The diagrams in  Fig. \ref{fig:nhu} indicate that \oi\ emission is possible for the ionization parameter associated to the emitting gas of the VBC, $\log U \sim -1$ \citep{marzianietal10}, where \caii\ emission should be negligible. The implication is that \oi\ and \caii\ should show different profiles, especially in Pop. B sources. The \oi\ + \caii\ blend in Pop. B sources indeed shows a striking similarity to the \hb\ profile, and we have considered a very broad component along with a broad component in the fit of Fig. \ref{fig:analysis}. 

\section{Conclusion}

The aim of this paper was not to conduct an extensive review on the formation of the LILs in the spectra of quasars, but rather to gather together pieces and lines of evidence under the framework of the 4DE1 with the hope of gaining additional constraints on the origin of the LILs. We can tentatively conclude that both \caii\ and \feii\ are mainly emitted in the same dense region at low ionization \citep{matsuokaetal08,bruhweilerverner08} since the values we derive from \caii\ are consistent with the ones of the best model for \feii\ obtained, for example, by \citet{bruhweilerverner08}.  This dense low ionization region appears to be present in both Pop. B and Pop.  A sources. In extreme Pop. A sources \citep{dultzinetal11,negreteetal12} it may be the only region left contributing to LIL emission, while for Pop. B and some (most?) Pop. A sources there is probably a wide gradient in density and physical properties. Nonetheless, if our results are correct, the difference in prominence of \feii\ emission may be associated more to the relative contribution of this dense region to the total line flux rather than to a continuous change in physical properties from Pop. A to Pop. B \citep{marzianietal10}. 

Conventional photoionization models can account for \rfe $\ltsim 0.5$\ if iron abundance is solar; the upper limit is increased to \rfe $\approx 1$\ if abundances are 5--10 times super solar. Sources that are strong \feii\ emitters  have \rfe$\gtsim 1$\ and are relatively rare, $\approx$ 10 \%\ in well-defined flux-limited samples \citep{zamfiretal10,marzianietal13}. Most of them have 1 $\ltsim$ \rfe$\ltsim 1.5$; only $\approx$ 2\%\ of all sources have \rfe$\gtsim 1.5$\ \citep{marzianietal13a}. Therefore, one can say that photoionization models come very close to accounting for \feii\ emission even for relatively rare strong \feii\ emitters. Nonetheless, it seems important that LIL emission in these extreme Pop. A sources is further investigated. They are the sources with the strongest evidence of a wind \citep{marzianietal96,leighlymoore04,marzianietal10}; shocks cannot be excluded as an excitation/ionization mechanism. Extreme Pop. A sources show very similar spectra in terms of diagnostic line ratios and profile shapes \citep{dultzinetal11,negreteetal12}. Similar structure and dynamics are probably at the origin of the spectral similarity, and it is at least conceivable that a strong wind transfers   mechanical energy to the outer regions of the accretion disk, at the same time shielding them from the central ionizing source \citep{leighly04,leighlymoore04}.

\begin{multicols}{2}\small
\bibliographystyle{apj} 

\begin{thebibliography}{69}
\expandafter\ifx\csname natexlab\endcsname\relax\def\natexlab#1{#1}\fi

\bibitem[{{Antonucci}(2013)}]{antonucci13}
{Antonucci}, R. 2013, Nature, 495, 165

\bibitem[{{Barth} {et~al.}(2013){Barth}, {Pancoast}, {Bennert}, {Brewer},
  {Canalizo}, {Filippenko}, {Gates}, {Greene}, {Li}, {Malkan}, {Sand}, {Stern},
  {Treu}, {Woo}, {Assef}, {Bae}, {Buehler}, {Cenko}, {Clubb}, {Cooper},
  {Diamond-Stanic}, {H{\"o}nig}, {Joner}, {Laney}, {Lazarova}, {Nierenberg},
  {Silverman}, {Tollerud}, \& {Walsh}}]{barthetal13}
{Barth}, A.~J., {et~al.} 2013, \apj, 769, 128

\bibitem[{{Boroson} \& {Green}(1992)}]{borosongreen92}
{Boroson}, T.~A., \& {Green}, R.~F. 1992, ApJS, 80, 109

\bibitem[{{Br\"uhweiler} \& {Verner}(2008)}]{bruhweilerverner08}
{Br\"uhweiler}, F., \& {Verner}, E. 2008, ApJ, 675, 83

\bibitem[{{Collin} \& {Joly}(2000)}]{collinjoly00}
{Collin}, S., \& {Joly}, M. 2000, NAR, 44, 531

\bibitem[{{Collin-Souffrin}(1987)}]{collinsouffrin87}
{Collin-Souffrin}, S. 1987, \aap, 179, 60

\bibitem[{{Collin-Souffrin} {et~al.}(1988){Collin-Souffrin}, {Dyson},
  {McDowell}, \& {Perry}}]{collinsouffrinetal88}
{Collin-Souffrin}, S., {Dyson}, J.~E., {McDowell}, J.~C., \& {Perry}, J.~J.
  1988, MNRAS, 232, 539

\bibitem[{{Collin-Souffrin} {et~al.}(1980){Collin-Souffrin}, {Joly}, {Dumont},
  \& {Heidmann}}]{collinsouffrinetal80}
{Collin-Souffrin}, S., {Joly}, M., {Dumont}, S., \& {Heidmann}, N. 1980, \aap,
  83, 190

\bibitem[{{Collin-Souffrin} {et~al.}(1986){Collin-Souffrin}, {Joly},
  {Pequignot}, \& {Dumont}}]{collinsouffrinetal86}
{Collin-Souffrin}, S., {Joly}, M., {Pequignot}, D., \& {Dumont}, S. 1986, \aap,
  166, 27

\bibitem[{{Davidson}(1972)}]{davidson72}
{Davidson}, K. 1972, \apj, 171, 213

\bibitem[{{Davidson} \& {Netzer}(1979)}]{davidsonnetzer79}
{Davidson}, K., \& {Netzer}, H. 1979, Reviews of Modern Physics, 51, 715

\bibitem[{{D'Onofrio} {et~al.}(2012){D'Onofrio}, {Marziani}, \& {
  Sulentic}}]{donofrioetal12}
{D'Onofrio}, M., {Marziani}, P., \& { Sulentic}, J.~W., eds. 2012, Astrophysics
  and Space Science Library, Vol. 386, Fifty Years of Quasars From Early
  Observations and Ideas to Future Research (Springer Verlag,
  Berlin-Heidelberg)

\bibitem[{{Dultzin} {et~al.}(2011){Dultzin}, {Martinez}, {Marziani},
  {Sulentic}, \& {Negrete}}]{dultzinetal11}
{Dultzin}, D., {Martinez}, M.~L., {Marziani}, P., {Sulentic}, J.~W., \&
  {Negrete}, A. 2011, in Proceedings of the conference ''Narrow-Line Seyfert 1
  Galaxies and their place in the Universe''. April 4-6, 2011. Milano, Italy.,
  ed. L.~F. et~al. (Eds.), Proceedings of Science

\bibitem[{{Dultzin-Hacyan} {et~al.}(1999){Dultzin-Hacyan}, {Taniguchi}, \&
  {Uranga}}]{dultzinhacyanetal99}
{Dultzin-Hacyan}, D., {Taniguchi}, Y., \& {Uranga}, L. 1999, in Astronomical
  Society of the Pacific Conference Series, Vol. 175, Structure and Kinematics
  of Quasar Broad Line Regions, ed. C.~M. {Gaskell}, W.~N. {Brandt},
  M.~{Dietrich}, D.~{Dultzin-Hacyan}, \& M.~{Eracleous}, 303

\bibitem[{{Dumont} {et~al.}(1998){Dumont}, {Collin-Souffrin}, \&
  {Nazarova}}]{dumontetal98}
{Dumont}, A.-M., {Collin-Souffrin}, S., \& {Nazarova}, L. 1998, \aap, 331, 11

\bibitem[{{Ferland} \& {Persson}(1989)}]{ferlandpersson89}
{Ferland}, G.~J., \& {Persson}, S.~E. 1989, \apj, 347, 656

\bibitem[{{Ferland} {et~al.}(2013){Ferland}, {Porter}, {van Hoof}, {Williams},
  {Abel}, {Lykins}, {Shaw}, {Henney}, \& {Stancil}}]{ferlandetal13}
{Ferland}, G.~J., {et~al.} 2013, RevMexA\&Ap, 49, 137

\bibitem[{{Garcia-Rissmann} {et~al.}(2012){Garcia-Rissmann},
  {Rodr{\'{\i}}guez-Ardila}, {Sigut}, \& {Pradhan}}]{garciarissmannetal12}
{Garcia-Rissmann}, A., {Rodr{\'{\i}}guez-Ardila}, A., {Sigut}, T.~A.~A., \&
  {Pradhan}, A.~K. 2012, \apj, 751, 7

\bibitem[{{Gaskell}(1982)}]{gaskell82}
{Gaskell}, C.~M. 1982, ApJ, 263, 79

\bibitem[{{Grandi}(1980)}]{grandi80}
{Grandi}, S.~A. 1980, \apj, 238, 10

\bibitem[{{Joly}(1987)}]{joly87}
{Joly}, M. 1987, \aap, 184, 33

\bibitem[{{Joly}(1989)}]{joly89}
---. 1989, \aap, 208, 47

\bibitem[{{Joly} {et~al.}(2008){Joly}, {V{\'e}ron-Cetty}, \&
  {V{\'e}ron}}]{jolyetal08}
{Joly}, M., {V{\'e}ron-Cetty}, M., \& {V{\'e}ron}, P. 2008, in Revista Mexicana
  de Astronomia y Astrofisica Conference Series, Vol.~32, Revista Mexicana de
  Astronomia y Astrofisica Conference Series, 59--61

\bibitem[{{Korista} \& {Goad}(2004)}]{koristagoad04}
{Korista}, K.~T., \& {Goad}, M.~R. 2004, ApJ, 606, 749

\bibitem[{{Landt} {et~al.}(2008){Landt}, {Bentz}, {Ward}, {Elvis}, {Peterson},
  {Korista}, \& {Karovska}}]{landtetal08}
{Landt}, H., {Bentz}, M.~C., {Ward}, M.~J., {Elvis}, M., {Peterson}, B.~M.,
  {Korista}, K.~T., \& {Karovska}, M. 2008, \apjs, 174, 282

\bibitem[{{Landt} {et~al.}(2011){Landt}, {Elvis}, {Ward}, {Bentz}, {Korista},
  \& {Karovska}}]{landtetal11}
{Landt}, H., {Elvis}, M., {Ward}, M.~J., {Bentz}, M.~C., {Korista}, K.~T., \&
  {Karovska}, M. 2011, \mnras, 414, 218

\bibitem[{{Leighly}(2004)}]{leighly04}
{Leighly}, K.~M. 2004, ApJ, 611, 125

\bibitem[{{Leighly} \& {Moore}(2004)}]{leighlymoore04}
{Leighly}, K.~M., \& {Moore}, J.~R. 2004, \apj, 611, 107

\bibitem[{{Marziani} {et~al.}(2006){Marziani}, {Dultzin-Hacyan}, \&
  {Sulentic}}]{marzianietal06}
{Marziani}, P., {Dultzin-Hacyan}, D., \& {Sulentic}, J.~W. 2006, {Accretion
  onto Supermassive Black Holes in Quasars: Learning from Optical/UV
  Observations} (New Developments in Black Hole Research), 123

\bibitem[{{Marziani} \& {Sulentic}(2012)}]{marzianisulentic12a}
{Marziani}, P., \& {Sulentic}, J.~W. 2012, The Astronomical Review, 7, 040000

\bibitem[{{Marziani} {et~al.}(2008){Marziani}, {Sulentic}, \&
  {Dultzin}}]{marzianietal08}
{Marziani}, P., {Sulentic}, J.~W., \& {Dultzin}, D. 2008, in Revista Mexicana
  de Astronomia y Astrofisica Conference Series, Vol.~32, Revista Mexicana de
  Astronomia y Astrofisica Conference Series, 69--73

\bibitem[{{Marziani} {et~al.}(1996){Marziani}, {Sulentic}, {Dultzin-Hacyan},
  {Calvani}, \& {Moles}}]{marzianietal96}
{Marziani}, P., {Sulentic}, J.~W., {Dultzin-Hacyan}, D., {Calvani}, M., \&
  {Moles}, M. 1996, ApJS, 104, 37

\bibitem[{{Marziani} {et~al.}(2010){Marziani}, {Sulentic}, {Negrete},
  {Dultzin}, {Zamfir}, \& {Bachev}}]{marzianietal10}
{Marziani}, P., {Sulentic}, J.~W., {Negrete}, C.~A., {Dultzin}, D., {Zamfir},
  S., \& {Bachev}, R. 2010, \mnras, 409, 1033

\bibitem[{{Marziani} {et~al.}(2013{\natexlab{a}}){Marziani}, {Sulentic},
  {Plauchu-Frayn}, \& {del Olmo}}]{marzianietal13a}
{Marziani}, P., {Sulentic}, J.~W., {Plauchu-Frayn}, I., \& {del Olmo}, A.
  2013{\natexlab{a}}, ArXiv e-prints

\bibitem[{{Marziani} {et~al.}(2013{\natexlab{b}}){Marziani}, {Sulentic},
  {Plauchu-Frayn}, \& {del Olmo}}]{marzianietal13}
---. 2013{\natexlab{b}}, ApJ, 764

\bibitem[{{Marziani} {et~al.}(2009){Marziani}, {Sulentic}, {Stirpe}, {Zamfir},
  \& {Calvani}}]{marzianietal09}
{Marziani}, P., {Sulentic}, J.~W., {Stirpe}, G.~M., {Zamfir}, S., \& {Calvani},
  M. 2009, A\&Ap, 495, 83

\bibitem[{{Marziani} {et~al.}(2003{\natexlab{a}}){Marziani}, {Sulentic},
  {Zamanov}, {Calvani}, {Dultzin-Hacyan}, {Bachev}, \&
  {Zwitter}}]{marzianietal03a}
{Marziani}, P., {Sulentic}, J.~W., {Zamanov}, R., {Calvani}, M.,
  {Dultzin-Hacyan}, D., {Bachev}, R., \& {Zwitter}, T. 2003{\natexlab{a}},
  ApJS, 145, 199

\bibitem[{{Marziani} {et~al.}(2001){Marziani}, {Sulentic}, {Zwitter},
  {Dultzin-Hacyan}, \& {Calvani}}]{marzianietal01}
{Marziani}, P., {Sulentic}, J.~W., {Zwitter}, T., {Dultzin-Hacyan}, D., \&
  {Calvani}, M. 2001, ApJ, 558, 553

\bibitem[{{Marziani} {et~al.}(2003{\natexlab{b}}){Marziani}, {Zamanov},
  {Sulentic}, \& {Calvani}}]{marzianietal03b}
{Marziani}, P., {Zamanov}, R.~K., {Sulentic}, J.~W., \& {Calvani}, M.
  2003{\natexlab{b}}, MNRAS, 345, 1133

\bibitem[{{Matsuoka} {et~al.}(2008){Matsuoka}, {Kawara}, \&
  {Oyabu}}]{matsuokaetal08}
{Matsuoka}, Y., {Kawara}, K., \& {Oyabu}, S. 2008, ApJ, 673, 62

\bibitem[{{Matsuoka} {et~al.}(2007){Matsuoka}, {Oyabu}, {Tsuzuki}, \&
  {Kawara}}]{matsuokaetal07}
{Matsuoka}, Y., {Oyabu}, S., {Tsuzuki}, Y., \& {Kawara}, K. 2007, \apj, 663,
  781

\bibitem[{{Negrete} {et~al.}(2012){Negrete}, {Dultzin}, {Marziani}, \&
  {Sulentic}}]{negreteetal12}
{Negrete}, A., {Dultzin}, D., {Marziani}, P., \& {Sulentic}, J. 2012, ApJ, 757,
  62

\bibitem[{{Netzer} \& {Marziani}(2010)}]{netzermarziani10}
{Netzer}, H., \& {Marziani}, P. 2010, \apj, 724, 318

\bibitem[{{Netzer} \& {Penston}(1976)}]{netzerpenston76}
{Netzer}, H., \& {Penston}, M.~V. 1976, \mnras, 174, 319

\bibitem[{{Netzer} \& {Wills}(1983)}]{netzerwills83}
{Netzer}, H., \& {Wills}, B.~J. 1983, \apj, 275, 445

\bibitem[{{Persson}(1988)}]{persson88}
{Persson}, S.~E. 1988, \apj, 330, 751

\bibitem[{{Peterson} {et~al.}(1982){Peterson}, {Foltz}, {Byard}, \&
  {Wagner}}]{petersonetal82}
{Peterson}, B.~M., {Foltz}, C.~B., {Byard}, P.~L., \& {Wagner}, R.~M. 1982,
  \apjs, 49, 469

\bibitem[{{Phillips}(1978)}]{phillips78b}
{Phillips}, M.~M. 1978, ApJ, 226, 736

\bibitem[{{Richards} {et~al.}(2006){Richards}, {Lacy}, {Storrie-Lombardi},
  {Hall}, {Gallagher}, {Hines}, {Fan}, {Papovich}, {Vanden Berk}, {Trammell},
  {Schneider}, {Vestergaard}, {York}, {Jester}, {Anderson}, {Budav{\'a}ri}, \&
  {Szalay}}]{richardsetal06}
{Richards}, G.~T., {et~al.} 2006, \apjs, 166, 470

\bibitem[{{Rodr{\'{\i}}guez-Ardila}
  {et~al.}(2002{\natexlab{a}}){Rodr{\'{\i}}guez-Ardila}, {Viegas}, {Pastoriza},
  \& {Prato}}]{rodriguezardilaetal02a}
{Rodr{\'{\i}}guez-Ardila}, A., {Viegas}, S.~M., {Pastoriza}, M.~G., \& {Prato},
  L. 2002{\natexlab{a}}, \apj, 565, 140

\bibitem[{{Rodr{\'{\i}}guez-Ardila}
  {et~al.}(2002{\natexlab{b}}){Rodr{\'{\i}}guez-Ardila}, {Viegas}, {Pastoriza},
  {Prato}, \& {Donzelli}}]{rodriguezardilaetal02b}
{Rodr{\'{\i}}guez-Ardila}, A., {Viegas}, S.~M., {Pastoriza}, M.~G., {Prato},
  L., \& {Donzelli}, C.~J. 2002{\natexlab{b}}, \apj, 572, 94

\bibitem[{{Shuder}(1981)}]{shuder81}
{Shuder}, J.~M. 1981, \apj, 244, 12

\bibitem[{{Sigut} \& {Pradhan}(1998)}]{sigutpradhan98}
{Sigut}, T.~A.~A., \& {Pradhan}, A.~K. 1998, \apjl, 499, L139

\bibitem[{{Sigut} \& {Pradhan}(2003)}]{sigutpradhan03}
---. 2003, ApJS, 145, 15

\bibitem[{{Sulentic} {et~al.}(2011){Sulentic}, {Marziani}, \&
  {Zamfir}}]{sulenticetal11}
{Sulentic}, J., {Marziani}, P., \& {Zamfir}, S. 2011, Baltic Astronomy, 20, 427

\bibitem[{{Sulentic} {et~al.}(2007){Sulentic}, {Bachev}, {Marziani}, {Negrete},
  \& {Dultzin}}]{sulenticetal07}
{Sulentic}, J.~W., {Bachev}, R., {Marziani}, P., {Negrete}, C.~A., \&
  {Dultzin}, D. 2007, ApJ, 666, 757

\bibitem[{{Sulentic} {et~al.}(2001){Sulentic}, {Marziani}, \&
  {Calvani}}]{sulenticetal01}
{Sulentic}, J.~W., {Marziani}, P., \& {Calvani}, M. 2001, X-ray Astronomy:
  Stellar Endpoints, AGN, and the Diffuse X-ray Background, 599, 963

\bibitem[{{Sulentic} {et~al.}(2000{\natexlab{a}}){Sulentic}, {Marziani}, \&
  {Dultzin-Hacyan}}]{sulenticetal00a}
{Sulentic}, J.~W., {Marziani}, P., \& {Dultzin-Hacyan}, D. 2000{\natexlab{a}},
  ARA\&A, 38, 521

\bibitem[{{Sulentic} {et~al.}(2012){Sulentic}, {Marziani}, \&
  M.}]{sulenticetal12a}
{Sulentic}, J.~W., {Marziani}, P., \& M., D. 2012, Astrophysics and Space
  Science Library, Vol. 386, {Fifty Years of Quasars}, ed. M.~{D'Onofrio},
  P.~{Marziani}, \& J.~W. {Sulentic} (Springer Verlag, Berlin-Heidelberg)

\bibitem[{{Sulentic} {et~al.}(2000{\natexlab{b}}){Sulentic}, {Marziani},
  {Zwitter}, {Dultzin-Hacyan}, \& {Calvani}}]{sulenticetal00b}
{Sulentic}, J.~W., {Marziani}, P., {Zwitter}, T., {Dultzin-Hacyan}, D., \&
  {Calvani}, M. 2000{\natexlab{b}}, ApJL, 545, L15

\bibitem[{{Sulentic} {et~al.}(2006){Sulentic}, {Repetto}, {Stirpe}, {Marziani},
  {Dultzin-Hacyan}, \& {Calvani}}]{sulenticetal06}
{Sulentic}, J.~W., {Repetto}, P., {Stirpe}, G.~M., {Marziani}, P.,
  {Dultzin-Hacyan}, D., \& {Calvani}, M. 2006, A\&Ap, 456, 929

\bibitem[{{Sulentic} {et~al.}(2004){Sulentic}, {Stirpe}, {Marziani}, {Zamanov},
  {Calvani}, \& {Braito}}]{sulenticetal04}
{Sulentic}, J.~W., {Stirpe}, G.~M., {Marziani}, P., {Zamanov}, R., {Calvani},
  M., \& {Braito}, V. 2004, A\&Ap, 423, 121

\bibitem[{{Sulentic} {et~al.}(2003){Sulentic}, {Zamfir}, {Marziani}, {Bachev},
  {Calvani}, \& {Dultzin-Hacyan}}]{sulenticetal03}
{Sulentic}, J.~W., {Zamfir}, S., {Marziani}, P., {Bachev}, R., {Calvani}, M.,
  \& {Dultzin-Hacyan}, D. 2003, ApJL, 597, L17

\bibitem[{{Vanden Berk} {et~al.}(2001){Vanden Berk}, {Richards}, {Bauer},
  {Strauss}, {Schneider}, {Heckman}, {York}, {Hall}, \&
  {Fan}}]{vandenberketal01}
{Vanden Berk}, D.~E., {et~al.} 2001, AJ, 122, 549

\bibitem[{{Verner} {et~al.}(1999){Verner}, {Verner}, {Korista}, {Ferguson},
  {Hamann}, \& {Ferland}}]{verneretal99}
{Verner}, E.~M., {Verner}, D.~A., {Korista}, K.~T., {Ferguson}, J.~W.,
  {Hamann}, F., \& {Ferland}, G.~J. 1999, \apjs, 120, 101

\bibitem[{{Wills} {et~al.}(1985){Wills}, {Netzer}, \& {Wills}}]{willsetal85}
{Wills}, B.~J., {Netzer}, H., \& {Wills}, D. 1985, \apj, 288, 94

\bibitem[{{Zamanov} \& {Marziani}(2002)}]{zamanovmarziani02}
{Zamanov}, R., \& {Marziani}, P. 2002, ApJ, 571, L77

\bibitem[{{Zamfir} {et~al.}(2008){Zamfir}, {Sulentic}, \&
  {Marziani}}]{zamfiretal08}
{Zamfir}, S., {Sulentic}, J.~W., \& {Marziani}, P. 2008, MNRAS, 387, 856

\bibitem[{{Zamfir} {et~al.}(2010){Zamfir}, {Sulentic}, {Marziani}, \&
  {Dultzin}}]{zamfiretal10}
{Zamfir}, S., {Sulentic}, J.~W., {Marziani}, P., \& {Dultzin}, D. 2010, \mnras,
  403, 1759

\end{thebibliography}

\end{multicols}

\end{document}